\documentclass{iau}
\usepackage{graphicx}
\title[Linear Polarization as a Probe of Blazar Jets] %% [give here short title] %%
{Time-Variable Linear Polarization as a Probe of the Physical Conditions in the Compact Jets of Blazars}

\author[A. P. Marscher]   %% [give here the short author list; use "et al." if 3 authors or more] %%
{
Alan P. Marscher
}
\affiliation{
Institute for Astrophysical Research, Boston University \\ 
725 Commonwealth Ave., Boston, MA 02215 USA \\ 
email: {\tt marscher@bu.edu}
}

\pubyear{2014}
\volume{313} 
\pagerange{xx--xx}
\jname{Extragalactic jets from every angle}
\editors{F. Massaro, C.C. Cheung, E. Lopez, A. Siemiginowska, eds.}
\begin{document}

\maketitle

\begin{abstract}
A single measurement of linear polarization of a nonthermal source provides direct information about the mean direction and level of ordering of the magnetic field. Monitoring of the polarization in blazars, combined with millimeter-wave VLBI imaging in both total and polarized intensity, has the potential to determine the geometry of the magnetic field. This is a key probe of the physical processes in the relativistic jet, such as ordered field components, turbulence, magnetic reconnections, magnetic collimation and acceleration of the jet flow, particle acceleration, and radiative processes that produce extremely luminous, highly variable nonthermal emission. Well-sampled monitoring observations of multi-waveband flux and radio-optical polarization of blazars show a variety of behavior. In some cases, the observed polarization patterns appear systematic, while in others randomness dominates. Explanations involve helical magnetic fields, turbulence, and perhaps particle acceleration that depends on the angle between the magnetic field and shock fronts that might be present.
Simulations from the author's TEMZ model, with turbulent plasma crossing a standing conical shock in the jet,
show that a mixture of turbulent and toroidal magnetic field can produce the level of polarization
variability that is observed, even when the two field components are roughly equal.
\end{abstract}

\section{Introduction}

There are a number of questions about blazars that we hope to answer through combined observational
and theoretical efforts. How is the plasma in blazar jets accelerated to flow velocities
near the speed of light and focused to within $\lesssim 1^\circ$
\cite[(Jorstad et al. 2005]{J05}; \cite[Claussen-Brown et al. 2013)]{CB13}?
Where and how do extremely luminous outbursts and outrageously short flares of radiation occur? 
How are relativistic particles accelerated: by shocks, magnetic reconnections, turbulence, or some
other process? Possible answers to these questions usually involve magnetic fields. The currently
prevailing paradigm of jet launching, collimation, and acceleration requires a strong helical
magnetic field, at least within the inner parsec \cite[(e.g., Komissarov et al. 2007]{kom07};
\cite[Tchekhovskoy et al. 2011)]{Sasha11}. Acceleration of particles is thought to depend critically
on the geometry of the magnetic field \cite[(e.g., Summerlin \& Baring 2012)]{sb12}. And turbulent
magnetic fields can lead to second-order Fermi acceleration of particles, magnetic reconnections
\cite[(e.g., Kowal et al. 2012]{kow12}; \cite[Dexter et al. 2014)]{dex14},
and rapid flares \cite[(Marscher \& Jorstad 2010]{mj10}; \cite[Narayan \& Piran 2012]{np12};
\cite[Marscher 2014)]{M14}.

Fortunately, the magnetic field geometry directly affects an observable property of a blazar's
radiation: its polarization. We can therefore use observations of time-variable polarization at
millimeter to optical wavelengths and spatially resolved polarization on VLBI images to infer
the geometry of the field and its relation to the emission properties of blazars.

\section{Linear Polarization for Different Magnetic Field Configurations}

A favorite assumption of emission modelers is that the magnetic field can be approximated to
be completely tangled on all scales of interest, except when it is compressed by a shock wave
or some other phenomenon \cite[(e.g., Hughes, Aller, \& Aller 1989]{HAA89}; \cite[Cawthorne 2006)]{caw06}.
This would lead to zero linear polarization except where such compression occurs, and essentially
zero short-term fluctuations. Instead, the linear polarization of the synchrotron radiation
tends to be low --- a few to tens of percent ---relative to its value in a uniform field, but
non-zero, and it often fluctuates rapidly. A more realistic geometry consists of turbulent
cells. Consider the case of $N$ cells, each with a uniform but randomly directed magnetic field of
the same magnitude. The mean polarization is then $\langle \Pi \rangle = \Pi_{\rm max}N^{-1/2}$
\cite[(Burn 1966)]{Burn66}, where $\Pi_{\rm max}$ (a weak function of spectral index, usually 70--75\%
in an optically thin source) is the value in a uniform field. If turbulent cells are constantly
passing through the emission region, the degree of polarization fluctuates with a standard
deviation $\sigma(\Pi) \sim \langle \Pi \rangle^{1/2}$, while the electric-vector position angle
$\chi$ varies randomly, often executing apparent rotations that can exceed $180^\circ$. These
rotations in $\chi$ are usually quite irregular, but can sometimes be surprisingly smooth
\cite[(Jones 1988)]{jones88}.

Since a helical magnetic field is the main requirement of magnetic launching models of jets,
it may be the case that this geometry persists out to parsec scales. \cite[[See Gabuzda (2013)]{gab13}
and \cite[Gabuzda et al. (2014)]{gab14} for observational evidence in support of this.
On the other hand, current-driven
instabilities may eventually disrupt the helical ordering at end of the jet's
acceleration/collimation zone (ACZ) where the kinetic energy density reaches equipartition with
the magnetic energy density \cite[(e.g., Nalewajko \& Begelman 2012)]{NB12}.]
In these models, the helical field propagates down the jet with the plasma. The
degree of polarization depends on viewing angle $\theta$ and the bulk Lorentz factor $\Gamma$
\cite[(see Lyutikov et al. 2005)]{lyut05}. If $\theta=0^\circ$, the net linear polarization is
zero if the intensity is uniform across the jet, owing to symmetry. If the aberrated viewing angle
$\theta^\prime=90^\circ$ (which occurs when $\sin\;\theta = \Gamma^{-1}$), $\chi$ is in the direction
of the jet axis if $B_z^\prime > B_t^\prime$, and perpendicular to the axis if $B_t^\prime > B_z^\prime$.
The degree of polarization $\Pi$ depends on $B_t^\prime/B_z^\prime$. 
Other viewing angles yield polarization properties that are qualitatively similar to the side-on case.
Note that this dependence of $\chi$ on $\theta^\prime$ applies also to a field geometry that
corresponds to any superposition of toroidal and longitudinal field, of which a helical field is
a specific case. One could imagine, for example, that the longitudinal field consists of magnetic
loops that are stretched parallel to the jet axis by cross-jet velocity gradients
\cite[(e.g., Laing 1980)]{Laing80}.

Since nature tends to avoid ideal conditions, we should consider the case of a helical or
toroidal magnetic field
with a non-uniform intensity across a given cross-section of the jet. The polarization of the
area with the highest intensity will then determine the net polarization position angle $\chi$, while the
degree of polarization $\Pi$ can be quite low if the relative intensity enhancement is weak and tens of
percent if there is a particularly bright spot. Furthermore, if the bright spot --- which
presumably just has a higher
density of radiating particles than the rest of the cross-section --- is offset from the jet axis,
the corresponding parcel of plasma can execute a spiral trajectory about the axis
owing to rotation of the flow
that arises from rotation of the base anchored in the black hole's ergosphere or the inner accretion
disk \cite[(Vlahakis 2006)]{vlah06}. If the viewing angle to the jet axis is $0^\circ$, the observer
will see rotation of $\chi$ at a uniform rate \cite[(see Marscher 2013 for an illustration)]{M13}.
In the more common case when the viewing angle $\theta < \Gamma^{-1}$, (so that
$\theta^\prime\ll 90^\circ$), the rate of rotation of $\chi$ will vary smoothly and monotonically
during each turn; see \cite[Marscher et al. (2008,]{M08} \cite[2010)]{M10}, where the model
is applied to BL Lac and PKS~1510$-$089.

\section{Interpretation of Rotations of the Polarization Vector}

Rotations of the optical polarization vector in $\gamma$-ray bright blazars
appear to be quite common \cite[(see above and, e.g., Larionov et al. 2008]{Lar08}, \cite[2013a]{Lar13a},
\cite[2013b]{Lar13b}; \cite[Abdo et al. (2010)]{abdo10}; \cite[Kiehlmann et al. 2013]{K13};
\cite[Jorstad et al. 2013]{J13}; \cite[Aleksi\'c et al. 2014]{Alek14};
and \cite[Morozova et al. 2014)]{Mor14}. As discussed in the previous section, such events can
be explained by (1) a flow that is rotating through a helical magnetic field, (2) random walks of
a turbulently disordered field, or (3) a twisted jet. To this we add another possibility, proposed
by \cite[Zheng et al. (2014)]{Zhang14}: (4) the passage of a moving shock through a region with a
highly disordered field. The compression of the shock partially orders the field, but this ordering
is seen at different depths as time advances owing to light-travel delays, leading to an apparent
rotation of the polarization by as much as $180^\circ$ per shock.

When the position angle is not rotating, it generally fluctuates, often rapidly and sometimes wildly,
about its mean value (see the above references for examples). The degree of polarization tends to
do the same. This strongly implies that the
magnetic field is at least partially disordered, which is consistent with turbulence. Although
turbulence can also cause rotations of $\chi$, and therefore explain the main features of the
time variability of the polarization vector, the observed rotations are often much smoother than
expected from turbulence. In addition, the timing of the rotations often appears non-random, such
as just before the peak of a flare, contrary to the behavior of a strictly stochastic process. The
ultimate test of rotation caused by geometry or rotation of the flow in the jet is that the
rotation in a given blazar should always be in the same direction, clockwise or counterclockwise.
This seems to be the case for PKS~1510$-$089 \cite[(cf. the rotations reported by Marscher et al. 2010
with those in]{M10} \cite[Aleksi\'c et al. 2014)]{Alek14}.

\section{Turbulence in Blazar Jets}

Since the rapid fluctuations in linear polarization suggest the presence of turbulence, the author
\cite[(Marscher 2014)]{M14}
has been developing a numerical model (TEMZ --- Turbulent Extreme Multi-Zone) that attempts to
explain the multi-waveband flux and polarization variations of blazars. The key features of the
model include:\\
1. Turbulent ambient jet plasma, which accelerates electrons with a power-law energy
distribution through the second-order Fermi process and possibly magnetic reconnections. The turbulence
is realized in the model by dividing the jet into many cylindrical cells, the number of
which is selected to match the degree of polarization.\\
2. A conical standing shock that further accelerates electrons, with the amplification in energy
depending on the angle between the magnetic field of the turbulent cell and the shock normal
\cite[(e.g., Summerlin \& Baring 2012)]{sb12}. \cite[Cawthorne (2006)]{caw06} and
\cite[Cawthorne et al. (2013)]{caw13} have found that the polarization pattern of the ``core''
of some blazars, observed with the VLBA at 43 GHz, matches the predictions of turbulent
plasma compressed in a standing conical shock.\\
3. The dependence of the particle acceleration on magnetic field direction reduces the volume
filling factor of the emission at the highest frequencies for both synchrotron and inverse
Compton radiation. This in turn causes higher amplitude, shorter time-scale variations at the
higher frequencies. Since the number of turbulent cells $N(\nu)$ that radiate at higher frequencies
$\nu$ are more limited, the mean polarization also increases with frequency.

\begin{figure}[b]
% \vspace*{-2.0 cm}
\begin{center}
\includegraphics[width=8cm]{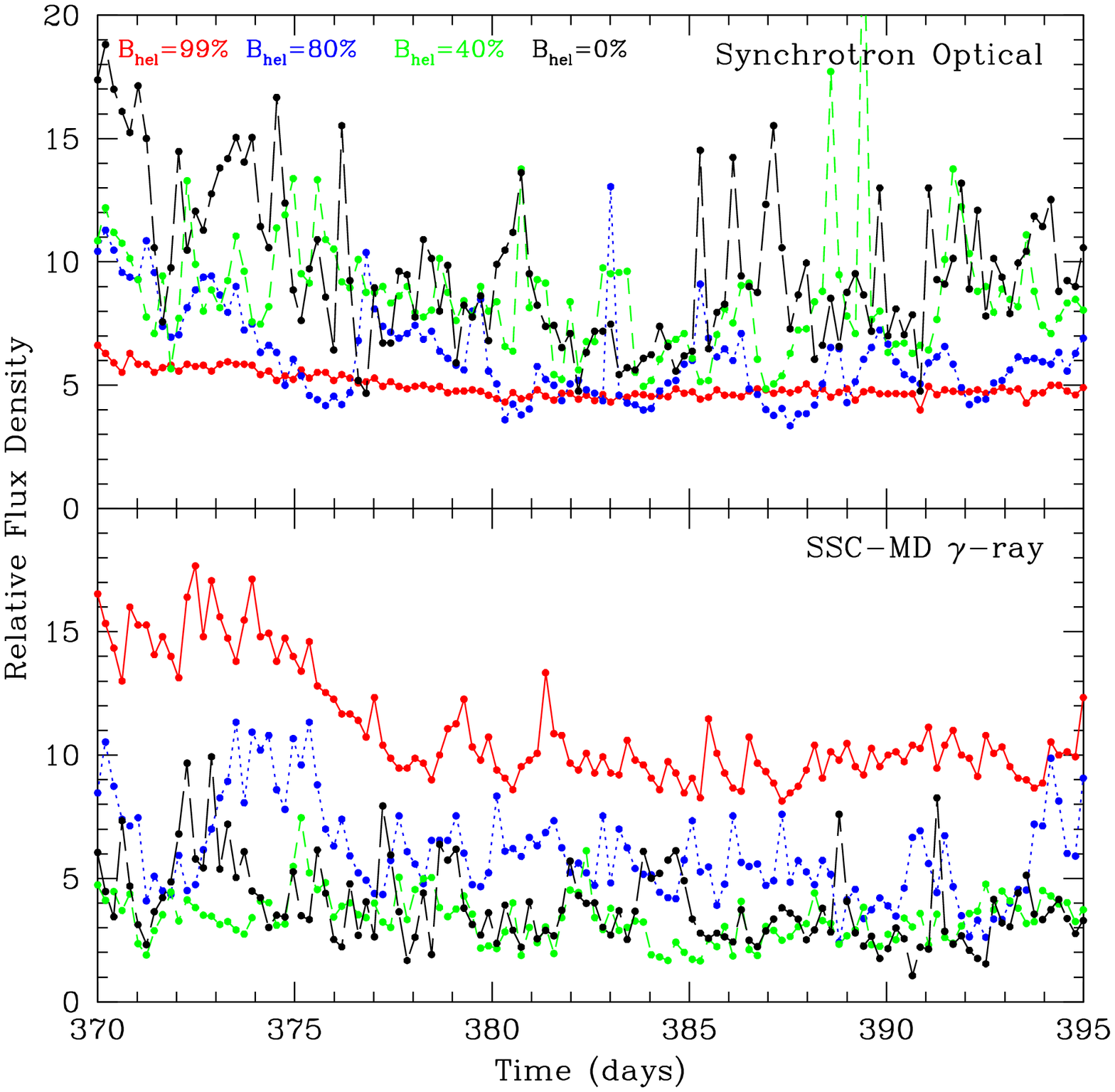} 
\vspace{8cm}
\includegraphics[width=8cm]{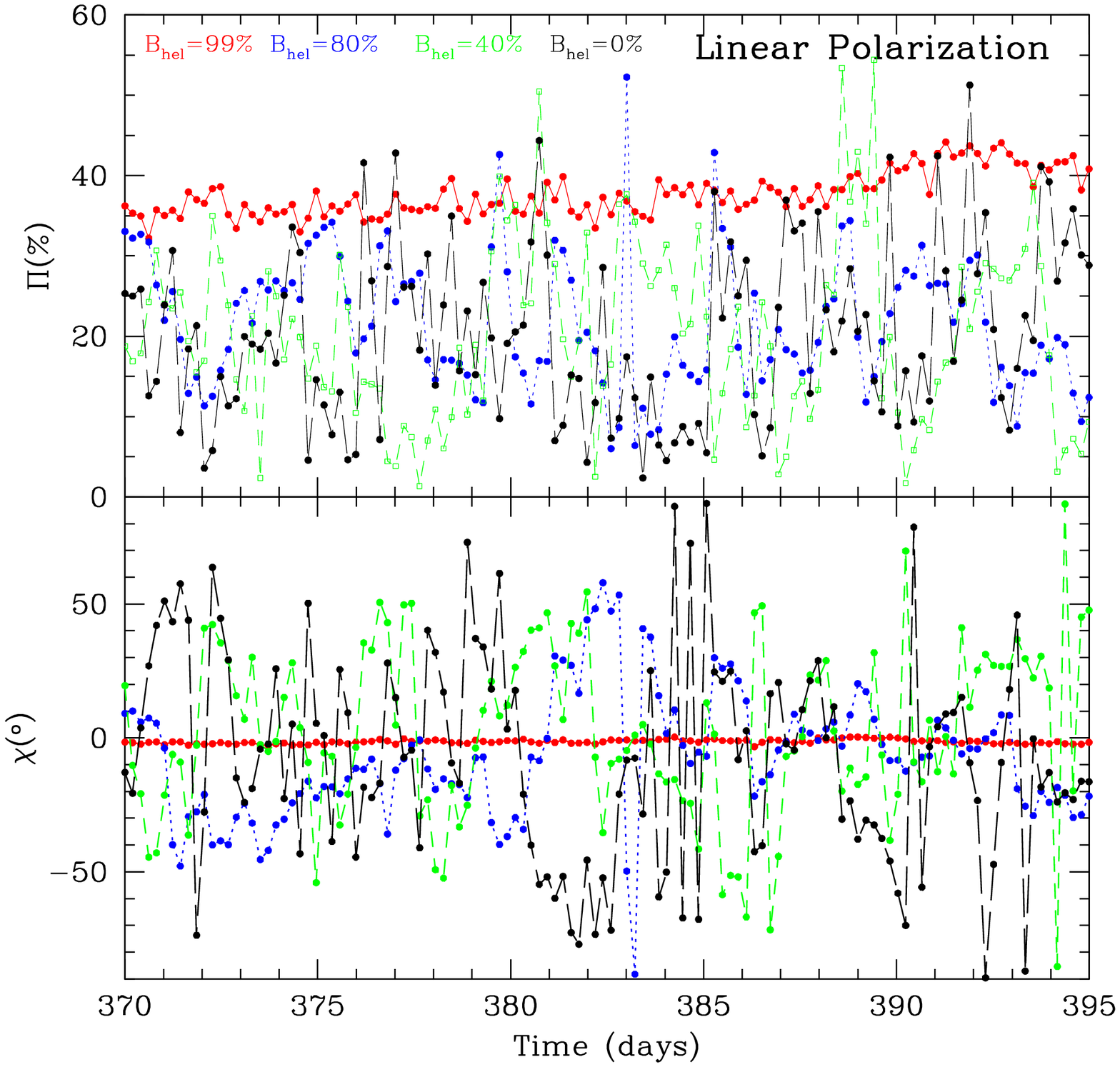} 
 \vspace*{-8.0 cm}
 \caption{\textit{Top:} Optical and $\gamma$-ray light curves during a 25-day time interval
from four runs of the TEMZ code
with different ratios of helical (pitch angle of $85^\circ$, hence nearly toroidal) to total
(helical + turbulent) magnetic field, as indicated (black \& white version --- solid: 0.99, dotted: 0.8, short-dashed: 0.4, long-dashed: 0). Parameters were selected to be similar
to the physical parameters of BL Lacertae. In this run, the seed photons for inverse Compton
scattering are synchrotron and synchrotron self-Compton (SSC) radiation emitted by relatively
slowly moving plasma in a Mach disk. \textit{Bottom:} Polarization vs.\ time for the same runs.}
   \label{fig1}
\end{center}
\end{figure}

\section{The Big Picture of a Blazar Jet}

A rough sketch of a blazar jet might then consist of a helical magnetic field in the
acceleration/collimation zone out to parsec scales, then turbulence (+ maybe magnetic reconnections)
dominates, perhaps alongside boundary layers where velocity shear stretches the magnetic field
in the longitudinal direction. Both moving shocks from major disturbances in the input energy
and/or velocity of the flow and standing shocks from pressure mismatches between the jet and
external medium, encounters with dense external gas after changes in jet direction, or collisions
with clouds, compress the magnetic field and further accelerate the radiating particles. This
general picture might be capable of producing the emission features that we see, including
rapid variations of flux and polarization out to parsec scales.

Since there is evidence that either helical or toroidal-plus-longitudinal magnetic fields
can be present on parsec scales, the question arises as to whether turbulent and helical fields
can co-exist in the same location. Since an ordered field should decrease the level of
variability below that observed, one might expect that the ordered component would need to be
a small fraction of the total field in blazars with rapidly variable polarization.
In order to test this, the author has run some TEMZ
simulations with various ratios of helical to total (helical + turbulent) field. The resulting
flux and polarization versus time curves are displayed in Figure \ref{fig1}. As can be seen,
the quenching of the variability is not apparent until the helical component composes considerably
more than 50\% of the total field. The conclusion is that less than 50\% of the field needs to
be disordered to explain --- qualitatively, at least --- the variability properties of blazars.
The author plans to use statistical tests to make the comparison between the model and data
more quantitative.

\section{Conclusions}

A combined international effort is now producing optical polarization data with sufficient time
coverage to follow variations in dozens of blazars. Even more data would be better, since
events such as rotations of the polarization vector are easy to miss when the sampling is sparse.
We are now identifying patterns in data --- some apparently systematic, others apparently random ---
that we can interpret in terms of physical properties of the jets. Further development of existing
and new theoretical models is needed to facilitate this. The author welcomes competition to his
own TEMZ model!

This research is supported in part by NASA through Fermi Guest Investigator grants NNX11AQ03G, NNX12AO79G, NNX13AP06G, and NNX14AQ58G.

\end{document}